\documentclass[prl,superscriptaddress]{revtex4}%
\usepackage{amsfonts}
\usepackage{amsmath}
\usepackage{amssymb}
\usepackage{graphicx}%
\providecommand{\U}[1]{\protect\rule{.1in}{.1in}}
\newcommand{\ket}[1]{ \ensuremath{\left| #1 \right\rangle} }
 \newcommand{\bra}[1]{ \ensuremath{\left\langle #1 \right|} }
 
 \newcommand{\IP}[2] { \ensuremath{\left\langle {#1} \left| {#2}\right.\right\rangle} }

\begin{document}

\title{On ``non-Hermitian Quantum Mechanics''.}
\author{Mark J. Everitt}
\email{m.j.everitt@physics.org}
\affiliation{Centre for Theoretical Physics, The British University in Egypt, El Sherouk City, Postal No. 11837, P.O. Box 43, Egypt.}
\affiliation{Department of Physics, Loughborough University, Loughborough, Leics LE11 3TU, United Kingdom}
\author{Shaaban Khalil}
\affiliation{Centre for Theoretical Physics, The British University in Egypt, El Sherouk City, Postal No. 11837, P.O. Box 43, Egypt.}
\author{Alexandre M. Zagoskin}
\affiliation{Department of Physics, Loughborough University, Loughborough, Leics LE11 3TU, United Kingdom}
\affiliation{Department of Physics and Astronomy, The University of British Columbia, Vancouver, Canada V6T 1Z1}

\begin{abstract}
\end{abstract}
\maketitle

A series of recent papers \emph{``Faster than Hermitian Quantum
Mechanics''}~\cite{b1} and related articles such as~\cite{b2,l1} made
a point of the possibility of a non-Hermitian, but PT-symmetric,
operator to play the role of a Hamiltonian. In particular, they
show that with an appropriate choice of an inner product, the
evolution generated by such an operator will conserve the norm and
scalar product. Before proceeding to our main  observation we
would first like to note that this choice of appropriate product
depends on the spectrum of the Hamiltonian. We feel that this is
problematic as it does not lend itself well to a general theory.
Furthermore, if one is to consider time dependant Hamiltonians,
this might lead to a theory where the inner product itself is not
only dependant on the Hamiltonian but also on time. Moreover, here
we would like to show that if one chooses such an inner product
then the Hamiltonian in question is actually Hermitian, and the
whole exercise is to a certain degree redundant.

An operator is termed Hermitian or self adjoint if $A^\dag =A$,
that is $\IP{Ax}{y}=\IP{x}{Ay}$ for all $x,y \in
\mathcal{H}$~\cite{Krey89,Dieu69}.
Therefore, Hermiticity of an operator is not an intrinsic property
of the operator itself, but rather a property of the operator with
reference to the Hilbert Space on whose elements it operates and,
in particular, the associated inner product.  For instance,
with conventional (dot) inner product, in the Heisenberg matrix formulation of quantum mechanics, $A^\dag$ is defined as
matrix transposition and complex conjugate, $A=(A^*)^T$, of the operator $A$. 
While the inner
product considered in~\cite{b1,b2} is not defined in this way and, in general, depends on the Hamiltonian itself. Hence $A^\dag$, and concomitantly Hermiticity, is defined with respect to this new inner product. The claim that $PT$-symmetric Hamiltonian is not
Hermitian because it does not satisfy the Hermiticity condition
with respect to the standard (dot) inner product is therefore not consistent with the choice of Hilbert space. As we will show
below, it must be a Hermitian (in the general sense of
Hermiticity) in order to have a consistent quantum mechanical
theory with conserved norm.

Let an operator $H$ generate the evolution through the Schr\"odinger equation,
$
i\hbar \frac{\partial }{\partial t}\ket{\psi} =H\ket{\psi}.
$
\ By multiplying on both sides by $\bra{\phi}$ we get
$
i\hbar \IP{\phi}{\frac{\partial }{\partial t}\psi} =\IP{\phi}{H\psi}.\label{eq:s1}
$
\ Similarly we find
$
i\hbar \IP{\psi}{\frac{\partial }{\partial t}\phi} =\IP{\psi}{H\phi}.\label{eq:s1bis}
$
\ Taking complex conjugate of the second equation and subtracting it from the first we find
$
i\hbar \left[ \IP{\phi}{\frac{\partial }{\partial t}\psi} +
\IP{\frac{\partial }{\partial t}\phi}{\psi}
\right]=\IP{\phi}{H\psi}-   \IP{H\phi}{\psi}.
    \label{eq:s4}
$
\ Now we recognise the left hand side as a total derivative with
respect to time, and use the definition of the Hermitian
conjugate, which yields
\begin{equation}
i\hbar  \frac{\partial }{\partial t} \IP{\phi}{\psi}
=
   \IP{\phi}{H\psi}-    \IP{\phi}{H^\dag \psi}  
=
   \IP{\phi}{\left(H-H^\dag \right)\psi}.
    \label{eq:s5}
\end{equation}
It follows from (\ref{eq:s5}) that if $H$ is Hermitian, then the
norm and the scalar product are conserved. On the other hand, if
the scalar product is conserved for any two functions from a full
basis of the Hilbert space, the Hamiltonian will be Hermitian.

In some textbooks such as~\cite{LandauV3}, normalisation together
with Schr\"odinger evolution is taken as a sufficient condition
that the Hamiltonian be Hermitian, which is not the case. The
preservation of the norm only fixes the diagonal matrix elements
of the operator $H$, $ \IP{\psi}{\left(H-H^\dag \right)\psi}$.


Hence, the structure of the Hilbert space imposed by introducing a
norm preserving inner product as introduced in~\cite{b1,b2}
implies the Hamiltonian is actually Hermitian. An interesting
corollary of this is that by changing the inner product on a
Hilbert space we also change the set comprising all Hermitian
operators acting on this space. As only members of this set can
represent observable we note that the ramifications of changing
the inner product extend beyond simply determining the Hermiticity
of the Hamiltonian.  In this respect, one concludes that what
is called as $PT$ symmetric quantum mechanics is just the ordinary
quantum mechanic with different (more involved and possibly problematic) inner product.
Therefore, this type of analysis can not be considered as a non-Hermitian
extension or generalisation of quantum mechanics.

\end{document}